# Would Russian solar energy projects be feasible independent of state subsidies?

Gordon Rausser[1], Galina Chebotareva[2], Wadim Strielkowski[1, 3*], Luboš Smutka[3]

[1] Department of Agricultural and Resource Economics, University of California, Giannini Hall, Berkeley, CA 94720, USA; rausser@berkeley.edu; strielkowski@berkeley.edu
[2] Department of Energy and Industrial Management Systems, Ural Federal University, Mira str. 19, 620002 Yekaterinburg, Russian Federation; g.s.chebotareva@urfu.ru
[3] Department of Trade and Finance, Faculty of Economics and Management, Czech University of Life Sciences Prague, Kamýcká 129, Prague 6, 165 00 Prague, Czech Republic; strielkowski@pef.czu.cz; smutka@pef.czu.cz

* Corresponding author: strielkowski@berkeley.edu

## Abstract

This paper explores the critical question of the sustainability of Russian solar energy initiatives in the absence of governmental financial support. The study aims to determine if Russian energy companies can maintain operations in the solar energy sector without relying on direct state subsidies. Methodologically, the analysis utilizes established investment metrics such as Net Present Value (NPV), Internal Rate of Return (IRR), and Discounted Payback Period (DPP), tailored to reflect the unique technical and economic aspects of Russian solar energy facilities and to evaluate the influence of sector-specific risks on project efficiency, using a rating approach. We examined eleven solar energy projects under ten different scenarios to understand the dynamics of direct state support, exploring variations in support cessation, reductions in financial assistance, and the projects' resilience to external risk factors. Our multi-criteria scenario assessment indicates that, under the prevailing market conditions, the Russian solar energy sector is not yet equipped to operate efficiently without ongoing state financial subsidies. Interestingly, our findings also suggest that the solar energy sector in Russia has a greater potential to reduce its dependence on state support compared to the wind energy sector. Based on these insights, we propose energy policy recommendations aimed at gradually minimizing direct government funding in the Russian renewable energy market. This strategy is designed to foster self-sufficiency and growth in the solar energy sector.

## 1. Introduction

Despite the recent economic and geopolitical events, Russia stands among the global leaders in traditional natural resource reserves and extraction, including natural gas, coal, and oil [1-6]. Despite the seemingly impractical development of alternative energy due to the low extraction costs and extended depletion periods of these resources, examples from countries like China and the United States demonstrate the feasibility of successfully developing renewable energy sources alongside traditional ones. Leading agencies' reports confirm that, despite increasing competition, these countries continue to lead in both investment and installed capacity in renewable energy sources as of early 2020 [7,8].

This study evaluates the potential for renewable energy deployment in Russia, challenging past characterizations of the country as merely "a gas station masquerading as a country" [9]. The authors assert the substantial potential for Russia to transition to a fossil-free economy, deeming it not only desirable but also economically viable [10-15]. We focus on the development of solar energy projects in Russia, in particular when the governmental financial support is concerned. Therefore, our main research question is the following one: "Would Russian solar energy projects be feasible without state subsidies, given the current economic and geopolitical circumstances, including Western sanctions?".

While the state currently supports renewable energy development in Russia, the authors of this study argue that the country possesses significant untapped potential in wind, solar, and small hydro energy. Figure 1 that follows demonstrates the total energy supply (TES) in Russia by source from 1990 until 2020.

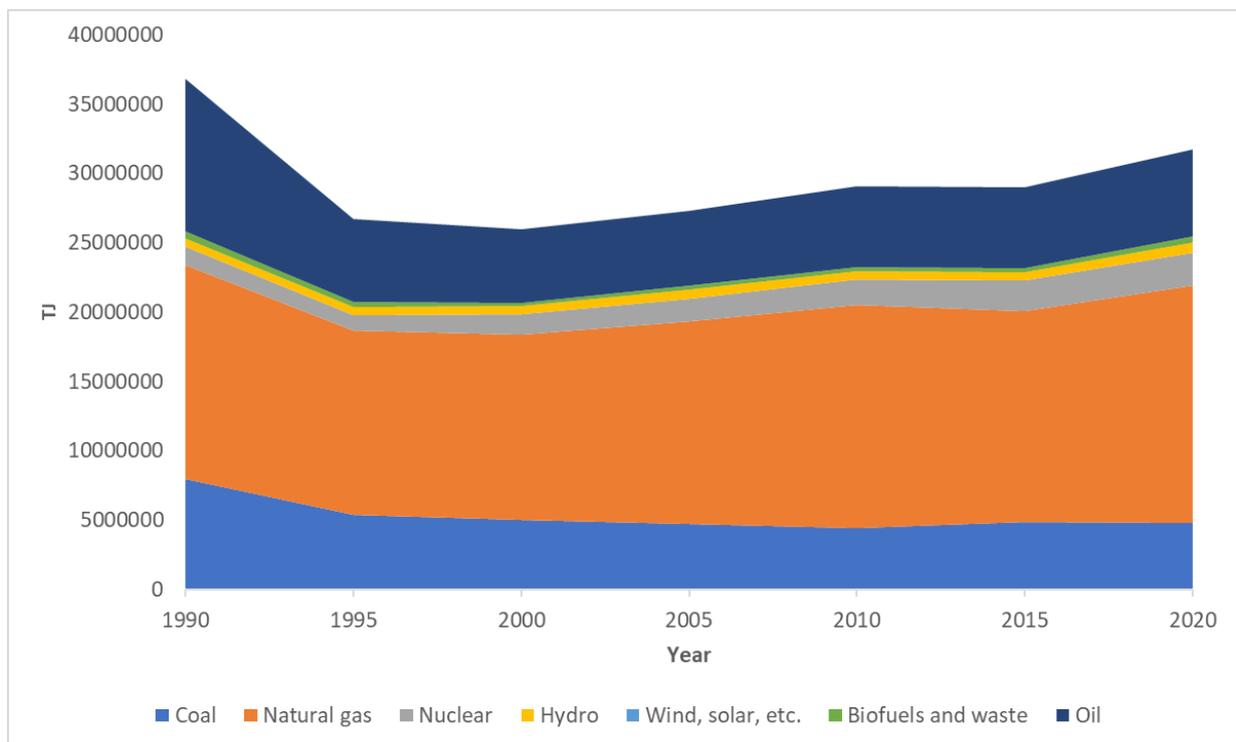

**Fig. 1.** Total energy supply in Russia by source (1990-2020).
Source: IEA (2023) [16]

Various incentive measures implemented at federal and regional levels include the selection of investment projects based on specific criteria, contracts for capacity supply, subsidies for technical connection costs, localization indicators, currency adjustment factors, mandatory purchase of renewable energy facilities by grid companies, establishment of long-term tariffs, and regional tax incentives. These measures collectively promote the growth of renewable energy projects in Russia [17, 18].

## 2. Solar energy market

Solar energy has conventionally stood out as one of the most rapidly expanding segments within the global renewable energy market [19]. As per the IEA report from 2023 [20], the expansion of installed capacity reached an unprecedented 243 GW in 2022. This achievement marked a record, surpassing the cumulative power level of 1 TW in solar power plants. Consequently, the proportion of electricity generated from solar sources rose from 5% to 6.2% by 2023. This notable advancement was predominantly attributed to the remarkable contributions of the top five countries—China, the USA, India, Brazil, and Spain—accounting for approximately 66% of the new installed capacity [21].

In order to replicate such achievements, global leaders have implemented various policy measures to bolster solar energy. For instance, in China, 93% of the rise in solar energy generation is linked to the three-year whole-county rooftop solar scheme initiated in early 2021. This initiative ensured the rapid development of distributed solar energy in the country [22]. The growth in Brazil is propelled by new regulations governing distributed generation. At the beginning of 2022, the government endorsed the use of photovoltaic systems of up to 5 MW for grid electricity accounting until 2045 and established grid connection fees starting in 2023 [23]. In the Spanish market, unsubsidized Power Purchase Agreements (PPAs) incentivize utility-scale PV systems, while high electricity prices drive rooftop PV systems [24].

Leading countries in the European solar energy market are also introducing supportive state measures for the sector. Norway has augmented the maximum subsidy for each 1 kW installed [25]. Germany has introduced tax incentives for residential and small PV systems [26]. Belgium has reduced taxes on photovoltaic installations and heat pumps in buildings less than ten years old [27]. Italy has eased permits for utility-scale PV systems and simplified the permitting process for commercial rooftop systems [28]. Portugal has streamlined permitting to promote self-consumption [29]. Austria has increased the budget for its rooftop solar rebate program [30].

For the subsequent advancement of solar energy, countries are continually formulating special programs. In the United States, despite the slowdown in new solar energy capacity commissioning in 2022, new measures were introduced in August [31]. These include an increase in the Investment Tax Credit from 26% to 30% for residential and commercial projects, and the approval of large-scale arrays to qualify for Production Tax Credits of up to 2.5 cents per kWh [32]. Japan has adopted a new mandate for solar PV, requiring all new homes and buildings to install rooftop PV starting in 2025 [33].

## 3. Materials and methods

Our research operates under the premise that solar energy projects in Russia can be economically viable without relying solely on direct state support, contingent upon their capacity to independently generate positive economic outcomes throughout the planned lifespan of the energy facilities. Accordingly, our analysis encompasses: 1) an examination of current mechanisms for incentivizing renewable energy projects prevalent in the global energy market; 2) the application of a multi-criteria step-by-step approach to assess the economic efficiency of solar energy projects, considering the prevailing economic conditions and sector-specific risks; 3) an evaluation of the economic viability of solar energy projects based on initial conditions; 4) the development of hypotheses and scenarios to explore the necessity of state support for energy projects based on the obtained results; 5) the testing of proposed hypotheses through a scenario-based evaluation of project effectiveness; and ultimately, 6) the formulation of recommendations to enhance state policies supporting solar energy in Russia.

The research relies on our proprietary methodology (e.g. Chebotareva et al., 2023) [34], originally designed for Russian renewable energy projects. In this study, the approach has been adapted for the multi-criteria assessment of investment indicators such as NPV, IRR, and DPP specific to solar energy projects. The alignment with multi-criteria assessment and industry nuances is achieved by considering the following factors:

- The intricacies in forming power prices for solar generation facilities on the wholesale Russian energy market (Government, 2022);
- Costs associated with the specific risks of Russian solar energy projects;
- Developed scenarios for state support of solar energy projects in Russia.

Figure 2 that follows provides a visual representation of the methodology employed for the ongoing examination of solar energy projects. Calculations are conducted across four defined phases of solar energy projects, with the phased assessment aimed at exploring the temporal capabilities of projects to achieve positive economic outcomes under varying scenario conditions, following the proposed algorithm (see Figure 2):

- 1st stage: Decision-making for project implementation;
- 2nd stage: Pre-operation of the facility;
- 3rd stage: Post-operation of the facility while maintaining state support, typically until the end of the 15-year support period;
- 4th stage: During the operation of the facility upon termination of support, usually until the end of the 25-year full planned life of the solar station.

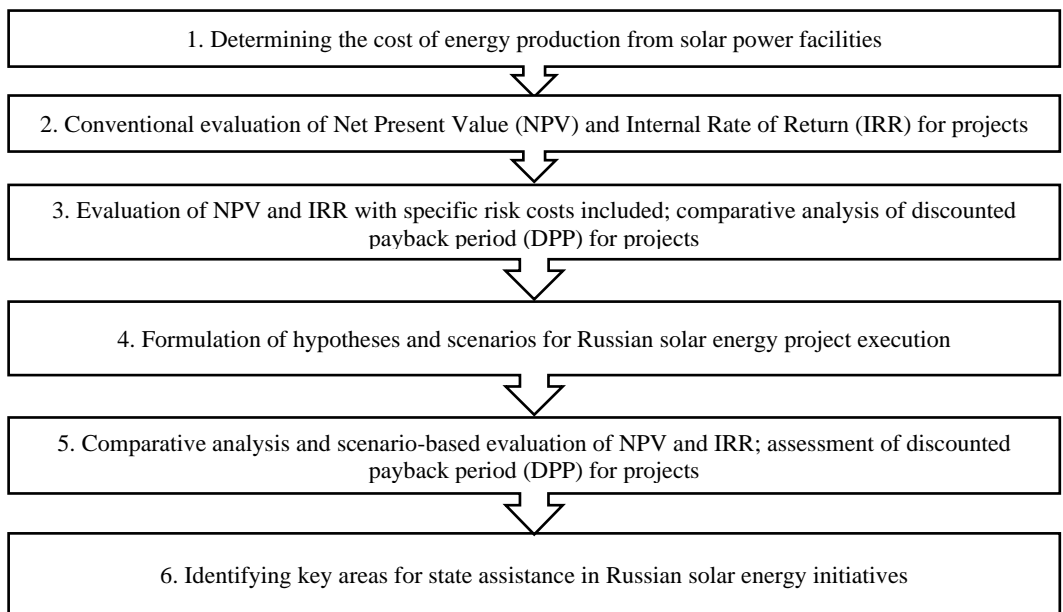

**Fig. 2.** Diagram showing the research methodology

## 4. Results and discussion

We conducted an evaluation of eleven solar energy projects currently underway in the wholesale sector of the Russian electricity market. The selection of these projects was based on the outcomes of the CSA RES competition [35]. A brief description of these projects together with their technical and economic characteristics is presented in greater detail in Table 1 that follows.

These initiatives are distributed across eight regions of Russia and fall within the first and second price zones, distinguished by their approaches to pricing for electrical energy and power formation. The projects are spearheaded by PJSC Forward Energo (formerly known as PJSC Fortum, Moscow), a prominent provider of electrical and thermal energy in the Urals and Western Siberia, and a key player in Russian renewable energy [36]. Additionally, LLC Avelar

Solar Technology (Moscow), a subsidiary of the Hevel Group, a trailblazer in Russian solar energy [37], is involved in the initiative.

**Table 1.** Main features of the studied Russian solar energy projects.

| Project name | Commissioning period, years* | Region/ price zone | Initiating company | Installed capacity** | Specific capital costs*** |
|---|---|---|---|---|---|
| SPP-2022-1 | 2,5 | Stavropol Territory / first | PJSC Forward Energo | 5,6 | 49,788 |
| SPP-2018-1 | 0,5 | Altai Republic / second | LLC Avelar Solar Technology | 10,0 | 122,000 |
| SPP-2018-2 | 0,5 | Altai Republic / second | LLC Avelar Solar Technology | 5,0 | 122,001 |
| SPP-2018-3 | 0,5 | Republic of Kalmykia / first | LLC Avelar Solar Technology | 23,5 | 122,002 |
| Astrakhan SPP | 2,5 | Astrakhan region / first | PJSC Forward Energo | 18,0 | 58,984 |
| Kalmykia SPP | 2,5 | Republic of Kalmykia / first | PJSC Forward Energo | 15,0 | 62,109 |
| Saratov SPP | 2,5 | Saratov region / first | PJSC Forward Energo | 15,0 | 62,805 |
| Orenburg SPP | 2,5 | Orenburg region / first | PJSC Forward Energo | 15,0 | 69,453 |
| Privolzhskaya SPP | 2,5 | Republic of Bashkortostan / first | PJSC Forward Energo | 15,0 | 69,853 |
| Privolzhskaya SPP-1 | 3,5 | Republic of Bashkortostan / first | PJSC Forward Energo | 17,0 | 58,901 |
| SPP Kalmykia | 3,5 | Republic of Kalmykia / first | PJSC Forward Energo | 15,0 | 59,103 |

Note: *calculated based on tender documentation; **MW; ***thous. RUB/kW.
Source: compiled from (ATS ENERGO, 2023)

Before going into the examination of the viability of state support for solar energy, we conducted a preliminary assessment of the economic efficiency of the projects. This evaluation, utilizing our research methodology, considered the costs associated with specific risks. Commentary on the Discounted Payback Period (DPP) criterion was provided in relation to conventional performance assessments that do not incorporate the costs of these risks.

The practical application of this approach across four stages revealed consistently negative effectiveness for the studied projects in the initial and subsequent stages. Consequently, these intermediate calculations were deemed inconsequential and were omitted from further consideration. Our results of the economic efficiency of the projects in question are presented in Table 2.

**Table 2.** Economic efficiency of projects (considering the cost of risks)

| Project name | Performance indicators | Third stage of the project* | The fourth stage of the project** | Author's comments on DPP*** (comparison with the classic assessment) |
|---|---|---|---|---|
| SPP-2022-1 | NPV, thous. RUB | **89 330,76** | 186 892,67 | Efficiency **is maintained** within the term of the RES CSA |
|  | IRR, % | **5,03** | 7,25 |  |
| SPP-2018-1 | NPV, thous. RUB | -415 823,51 | -303 504,55 | The project is initially **ineffective** |
|  | IRR, % | -6,37 | -3,51 |  |
| SPP-2018-2 | NPV, thous. RUB | -207 865,10 | -151 680,24 | The project is initially **ineffective** |
|  | IRR, % | -6,37 | -3,51 |  |
| SPP-2018-3 | NPV, thous. RUB | -808 078,09 | -453 495,85 | The project is initially **ineffective** |
| | IRR, % | -5,11 | -2,09 | |
| Astrakhan SPP | NPV, thous. RUB | **50 353,69** | 291 363,61 | Efficiency **is maintained** within the term of the RES CSA |
| | IRR, % | **0,83** | 3,38 | |
| Kalmykia SPP | NPV, thous. RUB | **59 012,94** | 289 505,29 | Efficiency **is maintained** within the term of the RES CSA |
| | IRR, % | **1,10** | 3,74 | |
| Saratov SPP | NPV, thous. RUB | -16 565,72 | **171 016,24** | Achievement of efficiency **is slowing down** before the planned service life of SPP |
| | IRR, % | -0,32 | **2,32** | |
| Orenburg SPP | NPV, thous. RUB | -51 057,52 | **155 463,95** | Achievement of efficiency **is slowing down** before the planned service life of SPP |
| | IRR, % | -0,89 | **1,91** | |
| Privolzhskaya SPP | NPV, thous. RUB | -74 349,76 | **120 076,20** | Efficiency **remains** within planned service life of SPP |
| | IRR, % | -1,30 | **1,50** | |
| Privolzhskaya SPP-1 | NPV, thous. RUB | -4 014,08 | **216 640,06** | Achievement of efficiency **is slowing down** before the planned service life of SPP |
| | IRR, % | -0,07 | **2,81** | |
| SPP Kalmykia | NPV, thous. RUB | **46 105,79** | 276 781,38 | Efficiency **is maintained** within the term of the RES CSA |
| | IRR, % | **0,94** | 3,85 | |

Note: *End of the CSA RES program; **End of the planned life of power plants; ***trends in the dynamics of project efficiency are highlighted in bold.

Source: own results

## 5. Conclusions

The imperative for the extensive implementation of renewable energy projects remains evident, even in countries abundant in fossil fuels, such as Russia. The adoption of renewable energy holds significance for nations aligning with the global environmental agenda and striving for energy security.

Our study investigated the prospect of the Russian solar energy sector abandoning direct state support for projects, specifically the priority preferential purchase of installed capacity on the wholesale market. To explore this, we formulated four hypotheses and devised ten scenarios modeling diverse economic conditions for project implementation. These scenarios ranged from amplifying the influence of specific risks to diminishing the volume, timing of financing, and even a complete cessation of state-supported subsidies.

The findings indicate that, under current circumstances, a complete withdrawal from direct state support is not economically viable for the Russian solar energy sector. In simulated conditions, considering subsidy schemes, market prices for electricity and capacity, and the adverse impact of sector-specific risks, projects fail to yield a positive economic outcome over the 25-year lifespan of power plants. Consequently, a nuanced examination of potential limits on reducing the volume and timing of state support for projects was necessitated.

Comparatively analyzing results from a parallel study on Russian wind energy projects (e.g., Chebotareva et al. (2023) study), it is noteworthy that solar energy deployment in Russia exhibits a slightly advanced position in terms of opportunities to curtail state support programs and funding compared to wind energy. This advancement stems from a more rapid reduction in the construction costs of solar stations and a shorter commissioning time (provided optimal unit costs are maintained). The identified six-month discrepancy in the proposed reduction of support periods (12.5 years for a solar power station versus 13 years for a wind power station) will significantly economize the budget, allowing for the selection of a greater number of renewable energy projects.

**Acknowledgements**

1. This research was supported by a grant of Russian Science Foundation (N 23-78-01242, https://rscf.ru/en/project/23-78-01242/) (sections 1, 2, 4.3, 5.1, 6).
2. This research was also partially supported by a grant from the Internal Grant Agency (IGA) of the Faculty of Economics and Management, Czech University of Life Sciences, project 2021B0002.